\documentstyle[aps,multicol,epsf]{revtex}
\draft
\begin{document}
\title{Atom Nano-lithography with Multi-layer Light Masks: \\ Particle Optics Analysis}
\author{R. Arun$^{1}$, I.Sh. Averbukh$^{1}$, and T. Pfau$^{2}$ }
\address{$^{1}$ Department of Chemical Physics, The Weizmann Institute of Science, 
Rehovot, Israel}
\address{$^{2}$ 5th Institute of Physics, University of Stuttgart, Germany}
\date{\today}
\maketitle
\begin{abstract}
We study the focusing of atoms by multiple layers of standing
light waves in the context of atom lithography. In particular,
atomic localization by a double-layer light mask is examined using
the optimal squeezing approach. Operation of the focusing setup is
analyzed both in the paraxial approximation and in the regime of
nonlinear spatial squeezing for the thin-thin as well as thin-thick 
atom lens combinations. It is shown that the optimized double light
mask may considerably reduce the imaging problems, improve the
quality of focusing and enhance the contrast ratio of the
deposited structures.
\end{abstract}

\begin{multicols}{2}
\section{introduction}
\vskip -0.15in
Since the early realization of sub-micron atom lithography
\cite{timp}, the subject of focusing neutral atoms by use of light
fields continues to attract a great deal of attention. The basic
principle of atom lithography relies on the possibility of
concentrating the atomic flux in space utilizing a spatially
modulated atom-light interaction. In the conventional
atom-lithographic schemes, a standing wave (SW) of light is used
as a mask on atoms to concentrate the atomic flux periodically and
create desired patterns at the nanometer scale \cite{review}. The
technique has been applied to many atomic-species in one
\cite{{prentis},{mcc1},{sodium},{mcc2},{chr},{alu},{ces},{ytt},{iro}} 
as well as two-dimensional \cite{twod} pattern formations. There are 
two ways to focus a parallel beam of atoms by light masks in close
correspondence with conventional optics. In the thin-lens
approach, atoms are focused outside the region of light field
which happens for low intensity light beams. On the other hand,
the atoms can be focused within the light beam when its intensity
is high. This is known as thick-lens regime and is very similar to
the graded-index lens of traditional optics. The laser focusing of
atoms depends on parameters such as thickness of light beam, velocity
spread of atoms, detuning of laser frequency from the atomic transition
frequencies, etc,. Experimentally, atomic nanostructures have been reported
with sodium \cite{{timp},{sodium}}, chromium \cite{{mcc2},{chr}},
aluminium \cite{alu}, cesium \cite{ces},
ytterbium \cite{ytt}, and iron \cite{iro} atoms.

Most of the theoretical studies on atom lithography employ a particle optics
approach to laser focusing of atoms \cite{{prentis},{mcc1},{ashkin}}.
The classical trajectories of atoms in the potentials induced by light fields suffice
to study the focal properties of light lens. In the case of direct
laser-guided atom deposition, the diffraction resolution limit will be ultimately
determined by the de Broglie wavelength of atoms, and may reach several picometers
for typical atomic beams \cite{lee}. In practice, however, this limit has never
been relevant because of the surface diffusion process, the quality of the atomic
beam, and severe aberrations due to anharmonicity of the sinusoidal dipole
potential. As a result, all current atom lithography schemes suffer from a considerable
background in the deposited structures. A possible way to overcome the aberration
problem was suggested in \cite{sch}, by using nanofabricated mechanical masks
that block atoms passing far from the minima of the dipole potential. However, this
complicates considerably the setup and reduces the deposition rate. Therefore, there
is a considerable need in a pure atom optics solution for the enhanced focusing of an
atomic beam having a significant angular spread.

In paraxial approximation, the steady-state propagation of an
atomic beam through a standing light wave is closely connected to
the problem of the time-dependent lateral motion of atoms subject
to a spatially periodic potential of an optical lattice. From this
point of view, enhanced focusing of the atomic beam can be
considered as a squeezing process on atoms in the optical lattice.
In  recent work \cite{mleib}, novel squeezing technique has been
introduced for atoms in a pulsed optical lattice. The approach
considered a time modulation of the SW with a series of short
laser pulses. Based on specially designed aperiodic sequence of
pulses, it has been shown that atoms can be squeezed to the minima
of the light-induced potential with reduced background level.
Oskay {\it et al.} \cite{raizen} have verified this proposal
experimentally using Cs atoms in an optical lattice. In Refs.
 \cite{{mleib},{raizen}}, the atoms were loaded into
  the optical lattice and the dynamics of atoms along the
direction of SW was studied as a time-dependent problem. The aim
of the present work is to extend the  focusing scenario of Ref.
\cite{mleib} to the beam configuration employed for atomic
nanofabrication. We generalize the results on atomic squeezing in
the pulsed SW to a system involving the atomic-beam traversing
several layers of light masks. In particular, we will investigate
prospects for reducing spherical and chromatic aberrations in atom
focusing with double-layer light masks. High-resolution deposition
of chromium atoms will be considered as an example. 

The plan of the paper is as follows. In Sec. II, the basic
framework of the problem is defined and the linear focusing of
atoms by a double-layer light mask is studied using the particle
optics approach in paraxial approximation. In Sec. III, we examine
the optimal squeezing scheme of \cite{mleib} in application to the
atomic-beam traversing two layers of light masks.  The effects of
beam collimation and chromatic aberrations are considered in Sec.
IV. Here, we optimize the double lens performance and give
parameters for the minimum spot-size in the atom deposition.
Finally, in Sec. V, we summarize our main results.

\section{squeezing of atoms by multi-layer light masks - classical treatment}
\vskip -0.15in
The focusing property of a single SW light has been studied in
great details by McClelland {\it et al.} \cite{{mcc1},{mcc2}}. The
light acts like an array of cylindrical lenses for the incident
atomic beam, focusing the atoms into a grating on the substrate.
However, because of the non-parabolic nature of the light-induced
potential, the focusing of atoms is subject to spherical
aberrations giving a finite width to the deposited features
\cite{mcc1}. A doublet of light masks made from two standing light
waves may, in principle, reduce the focusing imperfections due to
a clear physical mechanism. In this configuration, the first SW
prefocuses the atoms towards the minima of the sinusoidal
potential. When the pre-focused atoms cross the second SW, they
see closely the parabolic part of the potential which should
result in a reduction of the over-all spherical aberrations.

To test this scheme, we consider the propagation of an atomic-beam
through a combination of two SWs formed by counter-propagating
laser beams. The two SWs are identical except for their
intensities and are assumed to be formed along the x-direction.
Atoms are described as two-level systems with transition frequency
$\omega_o$. We take the direction of propagation of atoms through
the SW fields along the z-direction. If the atoms move
sufficiently slow (adiabatic conditions) through the light fields,
the internal variables of atoms maintain a steady state during
propagation \cite{cohen}. In this approximation, the atoms can be
described as point-like particles moving under the influence of an
average dipole-force. The potential energy of interaction is given
by \cite{{ashkin},{conserve}}
\begin{equation}
U(x,z) = \frac{\hbar \Delta}{2}~\hbox{ln}[1 + p(x,z)]~,
\label{poten}
\end{equation}
where
\begin{equation}
p(x,z) = \frac{\gamma^2}{\gamma^2 + 4 \Delta^2}~\frac{I(x,z)}{I_s}~.
\label{pxz}
\end{equation}
In Eq. (\ref{pxz}), $\Delta$ is the detuning of the laser frequency from
the atomic resonance, $I(x,z)$ is the light intensity, $\gamma$ is 
\vskip -0.2 in
\begin{figure}[t]
\epsfxsize=220pt
\centerline{
\epsfbox{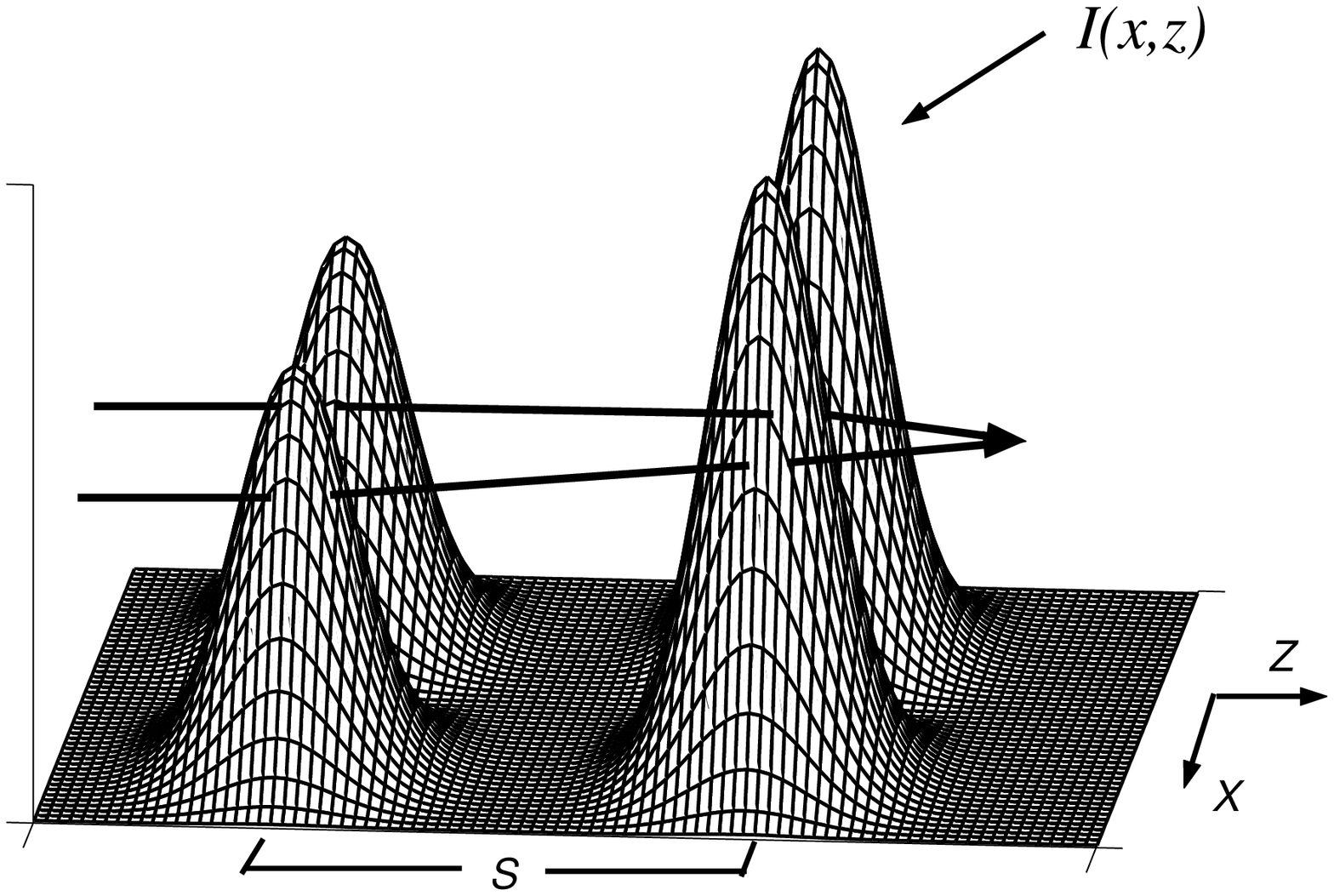}
}
\end{figure}
\vskip -0.1in
\noindent
FIG. 1. Schematic representation of the laser focusing of atoms by a
double layer of Gaussian standing waves. The intensity profile shows
the Gaussian envelopes along the z-axis and the sinusoidal variations along
the x-axis.
\vskip 0.2in
\noindent
the spontaneous decay rate of excited level, and $I_s$ is the saturation
intensity associated with the atomic transition. For the arrangement of
two SW light masks (denoted by 1 and 2) with separation $S$ between them,
the net intensity profile of light is given by
\begin{eqnarray}
I(x,z) &=& \left[I_1 \exp(-2 z^2/\sigma_z^2) + I_2 \exp(-2 {(z - S)}^{2}/\sigma_z^2)
\right] \nonumber \\
&&~~~~~ \times \sin^2(k x)~.
\end{eqnarray}
Here, $\sigma_z$ is the $1/e^2$ radius and $\lambda = 2 \pi/k$ is
the wavelength of laser beams forming the SWs. We  consider
Gaussian intensity profiles and ignore any y-dependence of laser
intensities as the force on atoms along the y-direction is
negligible compared to that along the direction of SW (x-axis).
$I_1$ and $I_2$ denote the maximum intensity of the standing light
waves 1 and 2, respectively. We neglect the overlap and
interference between two SWs. The intensity profile of light and
the focusing of atoms by light fields are shown schematically in
Fig. 1.

The classical trajectories of atoms in the potential
($\ref{poten}$) induced by the double-layer light masks obey the
Newton's equations of motions~:
\begin{equation}
\frac{d^2 x}{d t^2} + \frac{1}{m} \frac{\partial U(x,z)}{\partial x}
= 0~,~~~~
\frac{d^2 z}{d t^2} + \frac{1}{m} \frac{\partial U(x,z)}{\partial z} = 0~.
\end{equation}
Using the conservation of energy, we can combine the above two equations
and solve for $x$ as a function of $z$. This results in two first-order
coupled differential equations for $x(z)$, $\alpha \equiv dx(z)/dz$~~:
\begin{eqnarray}
\frac{dx(z)}{dz} &=& \alpha~~, \label{newton} \\
\frac{d\alpha(z)}{dz} &=& \frac{1 + \alpha^2}{2 (E - U)} \left(\alpha
\frac{dU}{dz} - (1 + \alpha^2) \frac{dU}{dx} \right)~~. \nonumber
\end{eqnarray}
Here, $E$ represents the total energy of the incoming atoms (the
kinetic energy in the field-free region) and $\alpha$ gives the
slope of the trajectory $x(z)$.

\vskip -0.2 in
\begin{figure}[t]
\centerline{ \epsfxsize=220 pt
\epsfbox{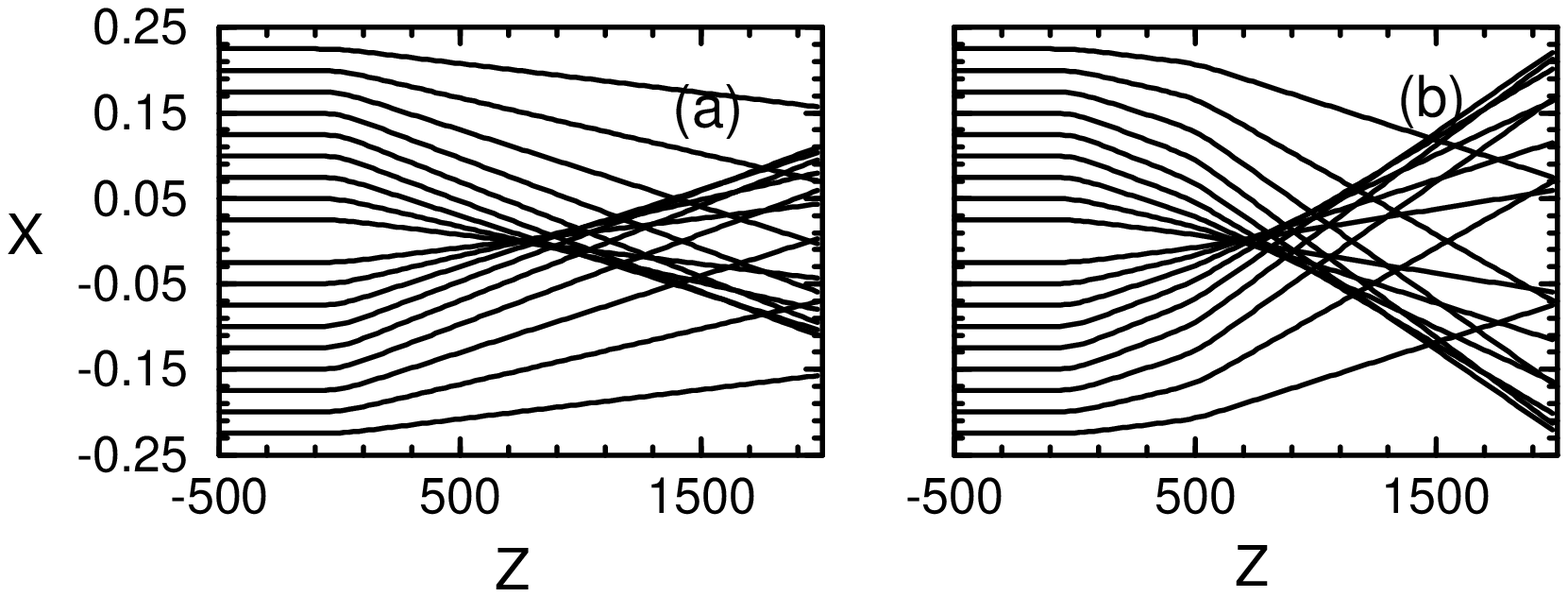}} 
\centerline{ \epsfxsize=220 pt
\epsfbox{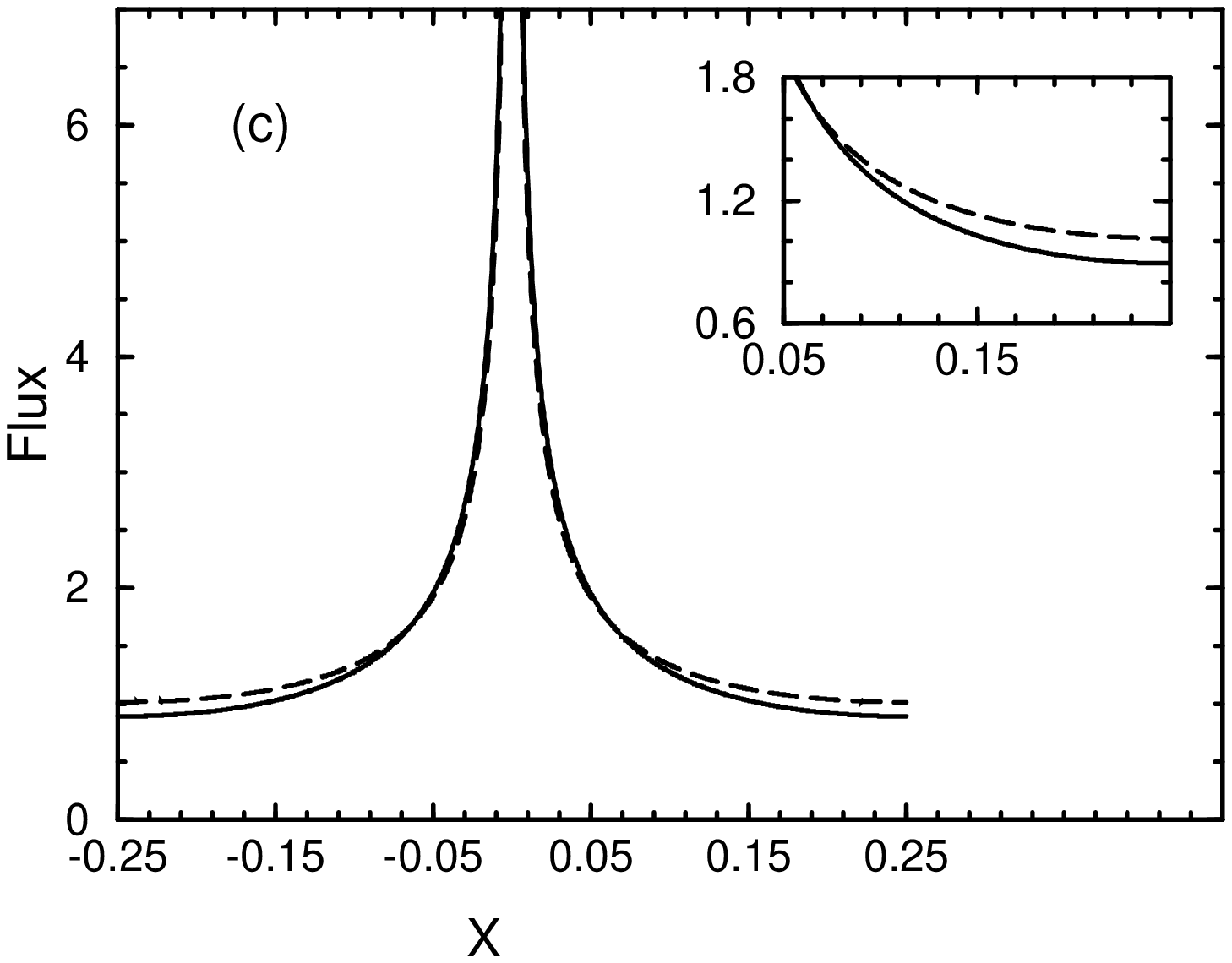}}
\end{figure}
\vskip -0.2in
\noindent
FIG. 2. Numerically calculated trajectories of atoms for laser focusing by 
a single- (a) and double-layer (b) light masks. The parameters used are
$I_1/I_s = 1000$, $I_2/I_s = 0$ (a) and $I_1/I_s = 1000$, $I_2 =
I_1$, $S = 500$ (b). All other parameters are the same as in Table
I. The solid (dashed) curve in graph (c) shows the probability
density of atoms at the focal point $z = z_f \approx 650~(700)$ of
the double (single) light lens. The region to the right of origin
$(x = 0)$ in graph (c) is zoomed and shown in the inset.
\vskip 0.2in

We first study the  focal properties of the light fields, and
solve numerically Eq. ($\ref{newton}$) for an atomic beam that is
initially parallel to the z-axis. The linear focal points and
principal-plane locations can be obtained by tracing paraxial
trajectories as discussed in \cite{mcc1}. Some typical results are
shown in Fig. 2, where we present the numerical calculation of a
series of atomic trajectories entering the nodal region of both
single $(I_1 \neq 0,I_2 \equiv 0)$ and double $(I_1,I_2 \neq 0)$
light masks. Table I lists the parameters used in dimensionless 
units, in which length is expressed in units of $\lambda$, and 
frequency is in units $\omega_{r} \equiv \hbar k^2 /2 m $ 
corresponding to the recoil energy. We have considered the 
intensities of light SWs to be equal in the case of double 
light masks. For the other variables, the values close to the 
experimental parameters of the chromium atom-deposition \cite{dirk} 
are taken as an example, though the general conclusions to be drawn 
should apply to other atoms. It is seen from Fig. 2, that a sharp 
focal spot appears in the flux of focused atoms \cite{pflux}. 
Despite the small size of the focal spot, the overall localization of 
atoms in the focal plane is not very marked. Atomic background in the
focal plane indeed gets reduced with double light masks as shown in the
inset of Fig. 2(c), however this effect is not very pronounced. To
take full advantage from double-mask arrangement, we have to replace the
concept of linear focusing (useful for paraxial trajectories only)
by the notion of optimized nonlinear spatial squeezing.

\section{optimal squeezing theory - \\application to atom nanolithography}
\vskip -0.1in
We have seen that the double light lens leads to some improvement
in feature contrast in the focal plane in comparison to the single
light lens. However, even for a single SW, the best squeezing of
atoms (maximal spatial compression) is achieved not at the focal
plane, but after the linear focusing phenomenon takes place. To
characterize the spatial localization of atoms we use a convenient
figure of merit, the localization factor \cite{mleib}:
\begin{eqnarray}
L(z) &=& 1 - <\cos\left(2 k x(z,x_o)\right)> \nonumber \\
&\equiv& \frac{2}{\lambda} \int_{-\lambda/4}^{\lambda/4} dx_o
\left[1 - \cos\left(2 k x(z,x_o)\right) \right]~, \label{local}
\end{eqnarray}
where $x(z,x_o)$ is the solution of the differential equations
($\ref{newton}$) satisfying the initial condition $x\rightarrow
x_{0}$  at $z \rightarrow - \infty$. The average in Eq.
($\ref{local}$) is taken over the random initial positions of
atoms and the localization factor is measured as a function of
distance $z$ from the center $(z=0)$ of the first SW. The
localization factor equals zero for an ideally localized atomic
ensemble, and is proportional to the mean-square variation of the
x coordinate (modulo standing wave period) in the case of 
well-localized distribution $(L << 1)$. 

\vskip 0.1in
Ref. \cite{mleib} considered the squeezing process in the
time-domain by analyzing the action of pulsed SWs on atoms. In the
Raman-Nath approximation, this corresponds to the thin-lens regime
(in space domain) for interaction of a propagating atomic-beam
with multiple layers of light masks. According to the optimal
squeezing strategy \cite{mleib}, the time sequence of pulses
applied to the atomic system is determined by minimizing the
localization factor. To apply this procedure to the atom squeezing
by multi-layer light masks, we should minimize the localization
factor ($\ref{local}$) in the parameter space: the separations
between the light SWs, their intensities, and the relative
distance of substrate surface with respect to the layers of light  
masks. This optimization can be done numerically 
\vskip 0.1in
\begin{table}[htb]
TABLE I. Parameters in scaled units. Frequency is measured in recoil
units, and length in units of the optical wavelength. Energy is given
in the units of recoil energy $\hbar \omega_r$.
\vskip 0.03in
\begin{tabular}[t]{lc}
~~~~~~~~~~Parameter & Numerical value \\ \hline
Spontaneous emission rate $\gamma$ & 238 \\
Detuning $\Delta$ & 9500 \\
1/$e^2$ radius of SW $\sigma_z$ & 120 \\
Energy of the incoming atoms $E$ & 3$~\times 10^9$  \\
\end{tabular}
\end{table}
\begin{figure}[t]
\epsfxsize=220pt
\centerline{
\epsfbox{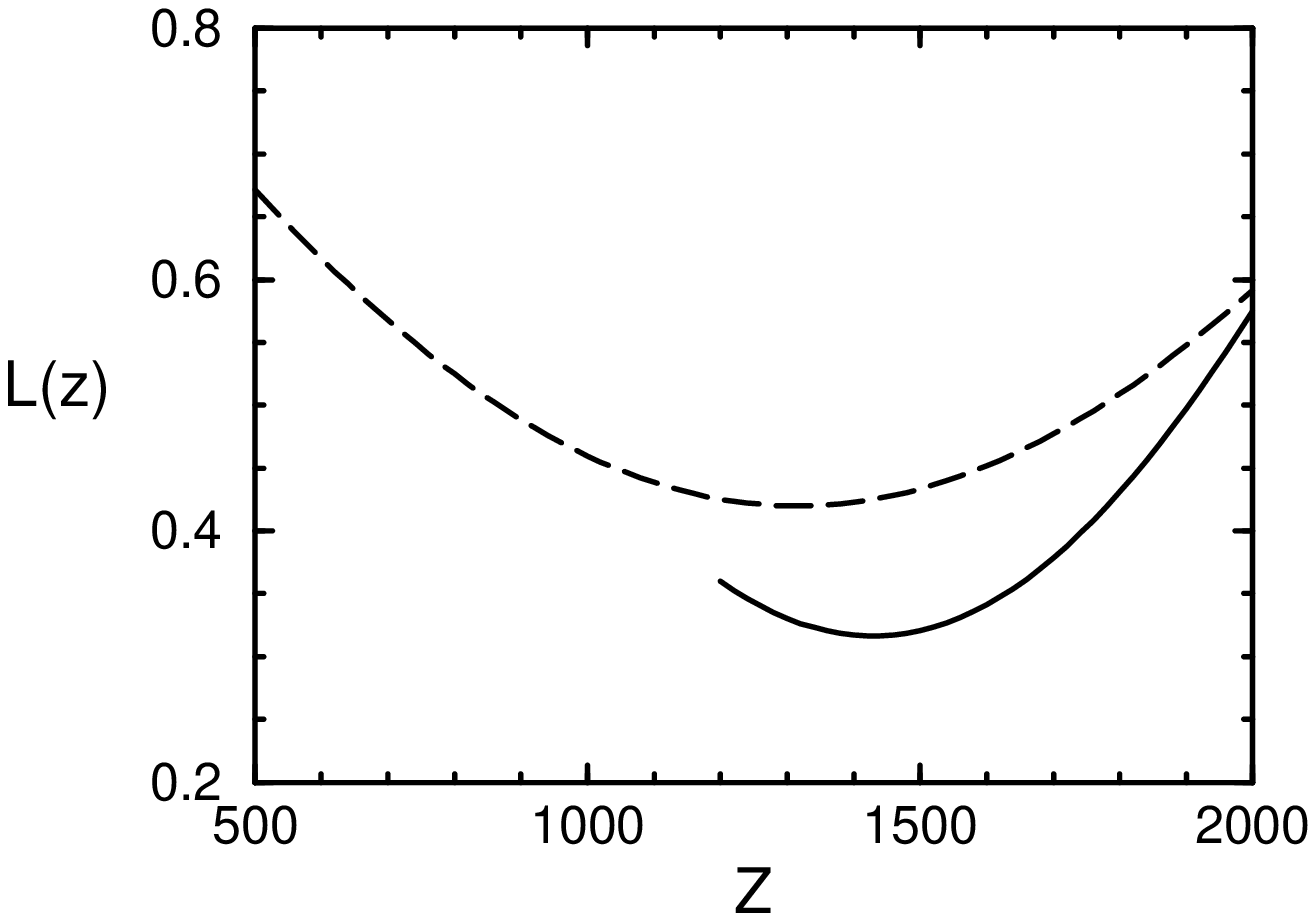}
}
\end{figure}
\vskip -0.2in
\noindent
FIG. 3. Localization factor of the atomic distribution for
squeezing by a single- (dashed curve) and double-layer (solid
curve) light masks. The parameters used are $I_1/I_s = 1000$,
$I_2/I_s = 0$ (dashed curve) and $I_1/I_s = 1000$, $I_2 = I_1$, $S
= S_m \approx 1000$ (solid curve). All other parameters are the
same as in Table I. The minimal value of $L(z)$ is 0.31 (0.42) and
it occurs at $z = z_m \approx 1450~(1300)$ for the solid (dashed)
curve. In the case of the double light mask, the point $(z_m,S_m)$
corresponds to the numerically found global minimum of the
localization factor.
\vskip 0.1in
\noindent
using the established simplex-search method. Our numerical analysis shows
that the localization factor exhibits multiple local minima even
for the simplest case of double light masks. In Fig. 3, we plot
the localization factor as a function of distance $z$ both for
single and double light masks around its global minimum
$(z_m,S_m)$. The intensities of SWs have been chosen to be equal
and satisfy the thin-lens condition of atom-light interaction
\cite{thin}. The graph shows that the localization factor gets a
sizable reduction with double light masks indicating for an
enhanced focusing of atoms. The minimum values of $L(z)$ in Fig. 3
are in conformity with the values obtained for optimal squeezing
of atoms with single and double pulses in the time-dependent
problem \cite{mleib}. We emphasize that the best squeezing
(localization) of atoms does not occur at the focal point. Figure
4 displays the spatial distribution of atoms at the point of best
localization. Instead of a single focal peak, a two-peaked spatial
distribution of atoms near the potential minima is observed in
Fig. 4. The origin of these peaks can be related to the formation
of rainbows in the wave optics and quantum mechanics, and it is
discussed in detail in \cite{{mleib},{rain}}. Moreover, on
comparing the inset of Figs. 2(c) and (4), it is seen that the
optimized separation between layers of the double light mask
results in a considerable reduction of atomic deposition  in the
background. This also leads to an overall increased concentration
of atomic flux near the potential minima.

We note, that according to \cite{{mleib},{raizen}}, further
squeezing of atoms can be achieved by increasing the number of
identical SWs in the multi-layer light masks. For the best
localization, again the optimized values for the separations
between light masks should be used.

\vskip 0.8in
\begin{figure}[h]
\epsfxsize=220pt
\centerline{
\epsfbox{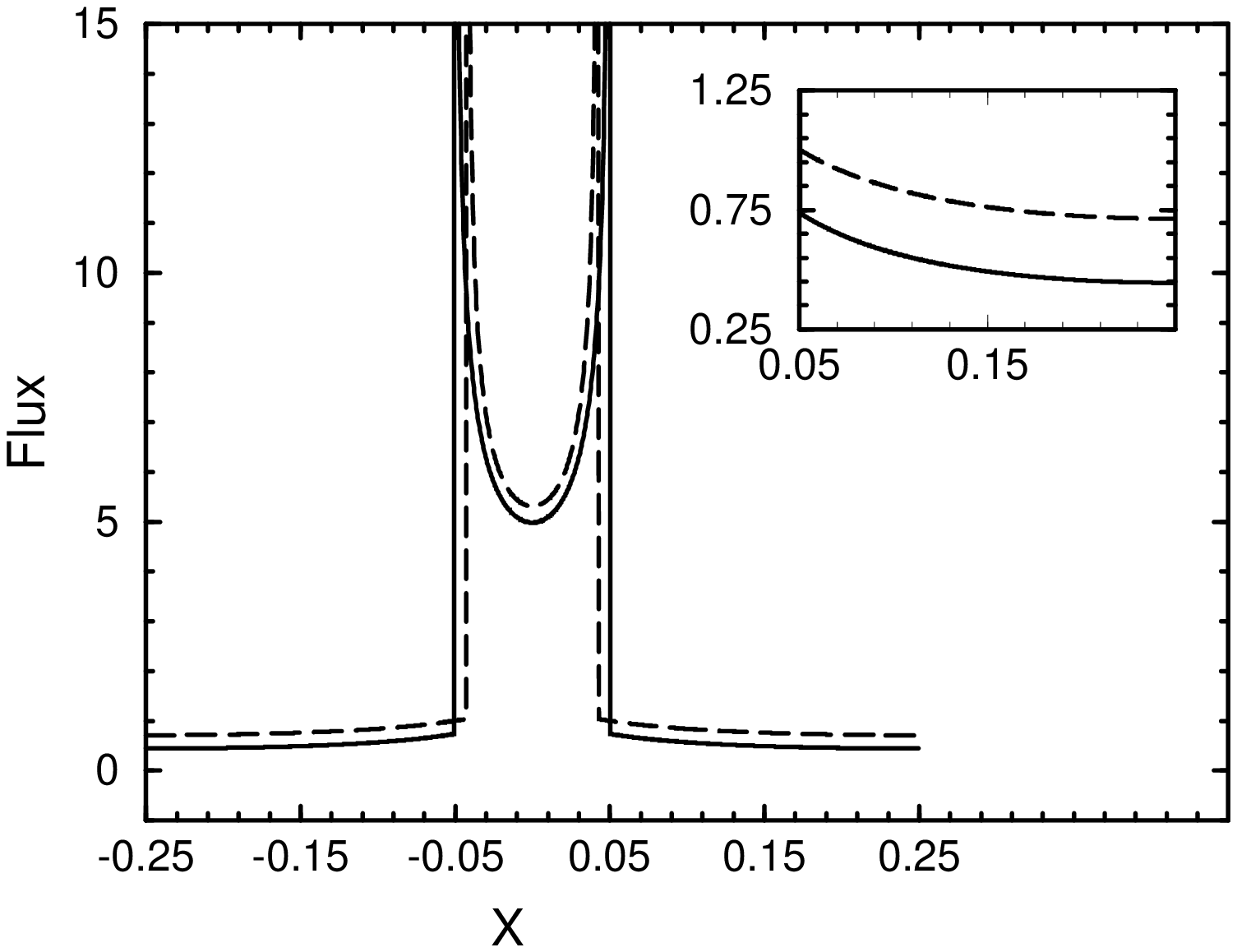}
}
\end{figure}
\vskip -0.5in
\noindent
FIG. 4. Probability density of atoms at the point of the best
squeezing by a single- (dashed curve) and double-layer (solid
curve) light masks. The parameters used are $I_1/I_s = 1000$,
$I_2/I_s = 0$, $z = z_m \approx 1300$ (dashed curve) and $I_1/I_s
= 1000$, $I_2 = I_1$, $S = S_m \approx 1000$, $z = z_m \approx
1450$ (solid curve). All other parameters are the same as in Table
I. The region to the right of origin $(x = 0)$ is zoomed and shown
in the inset.

\vskip -0.1in
In the above analysis, we have considered the case of equal
intensities for the light lenses and the problem has been studied
in the thin-lens \cite{firstthin} regime of atom focusing by light 
masks. However, in many current atom-lithographic schemes, focusing of 
atoms is generally achieved using an intense SW light. This corresponds 
to the thick-lens regime of atom-light interaction. In this limit,
the focal point is within or close to the region of laser fields and 
hence a detailed information on atomic motion within the light is 
required for a full description \cite{mcc1}. For the chromium atoms
deposition, the focusing of atoms to the center of an intense SW
has been extensively studied both theoretically \cite{mcc1} and
experimentally \cite{mcc2}. We show here that a combination of a
thin and thick lenses can result in the enhanced localization of
atoms 
\begin{figure}[h]
\epsfxsize=215pt
\centerline{
\epsfbox{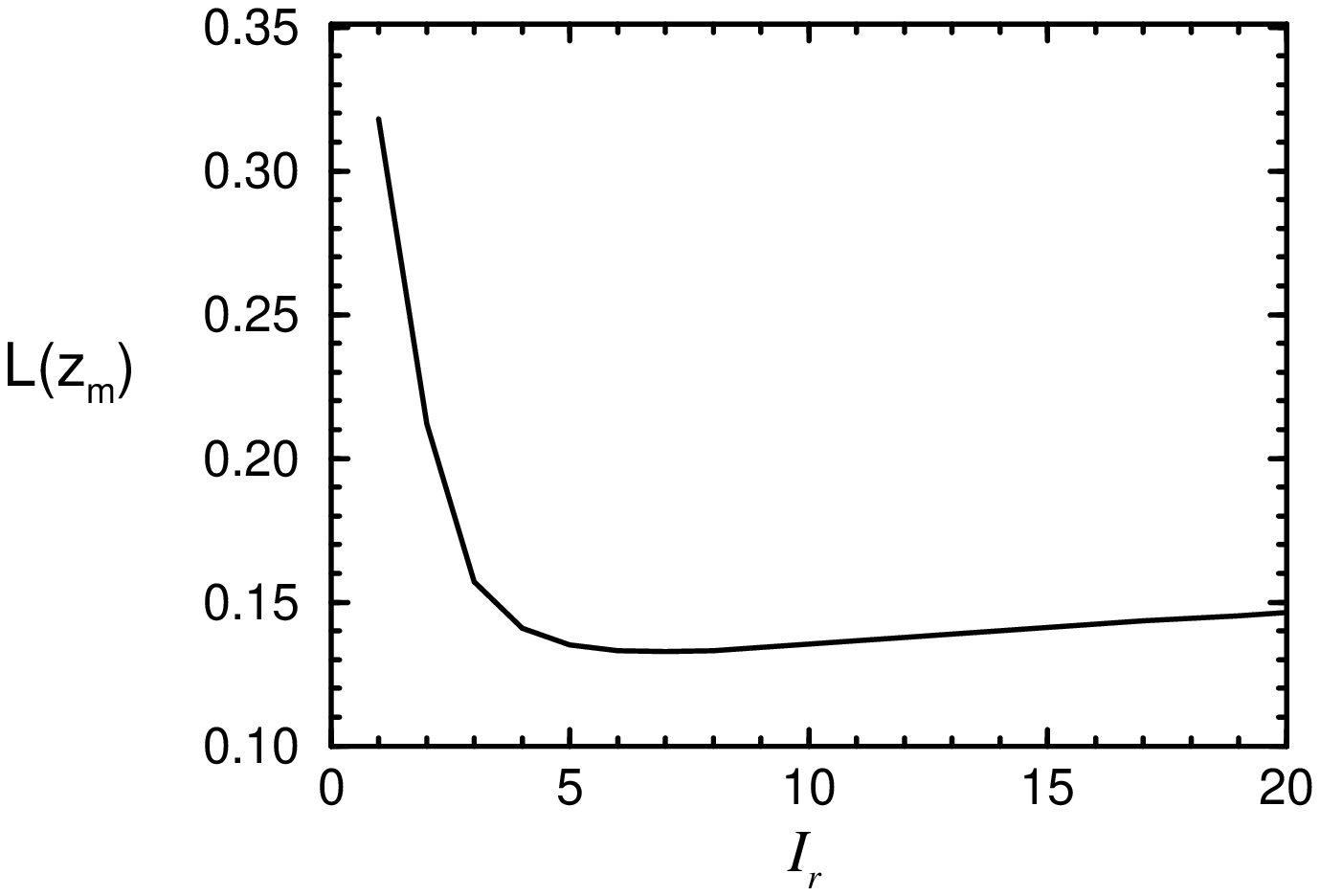}
}
\end{figure}
\vskip -0.5in
\noindent
FIG. 5. Minimal localization factor (maximal squeezing) of
the atomic distribution as a function of the relative intensity
$I_r \equiv I_2/I_1$ of standing light waves in a double light
mask. The parameters used are same as in Table I with $I_1/I_s =
1500$.

\noindent
with minimal background structures. For illustration, we
consider the focusing of atoms by a doublet of light masks made of
a thin and a thick lens. We fix the intensity of the first SW
light mask to satisfy the thin-lens limit and study the best
localization of atoms that can be achieved by varying the
intensity of the second SW. A plot of the minimal value of the
localization factor versus the relative intensity of the second
light mask is shown in Fig. 5. The graph shows that the
localization factor becomes almost insensitive to the variation in
relative intensity after the intensity ratio reaches the value of
5, and it approaches a small value of $L=0.15$. This result is to
be compared with the value of $L=0.31$ for the optimal squeezing
by two thin lenses, and $L = 0.42$ achievable by a single thin lens.
Fig. 6 shows the corresponding  trajectories of atoms and a plot of
atomic distribution at 

\begin{figure}[h]
\centerline{ \epsfxsize=185 pt \epsfbox{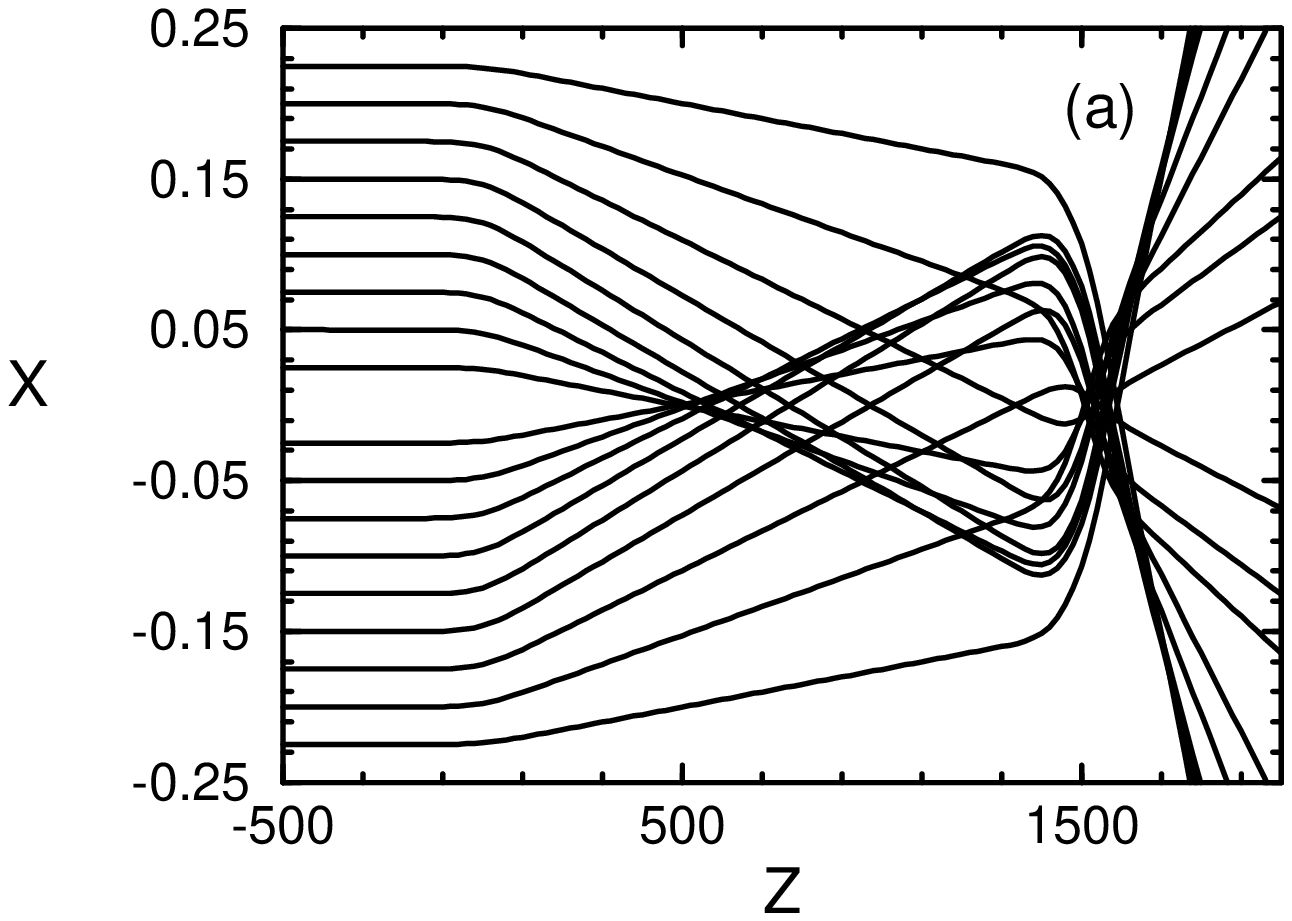}} \centerline{
\epsfxsize=215 pt \epsfbox{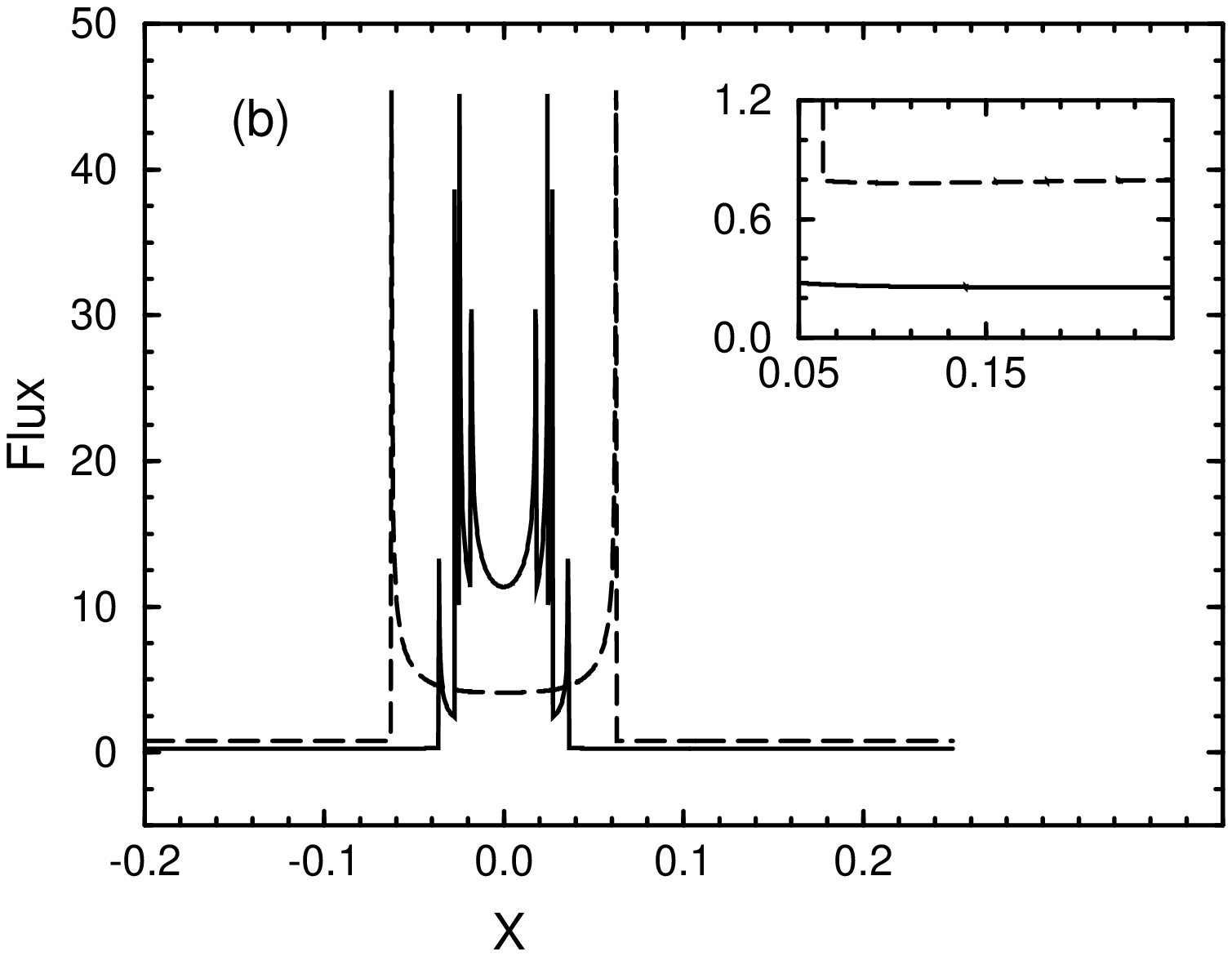}}
\end{figure}
\vskip -0.25in
\noindent
FIG. 6. (a) Numerical trajectory calculation for laser focusing
by a double light mask. The parameters used for the calculation
are $I_1/I_s = 1500$, $I_2 = 25 I_1$, and $S = S_m \approx 1500$.
All other parameters are the same as in Table I. (b) Probability
density of the atomic distribution at the point $(z_m,S_m)$ of maximal
squeezing by the double light mask. The parameters used are the
same as those of (a) with $z = z_m \approx 1550$. The point
$(z_m,S_m)$ is the numerically found global minimum of the
function $L(z)$ with respect to the variables $(z,S)$. The dashed
curve in graph (b) shows the atomic distribution at the point $z =
z_m$ of the best squeezing by a single thick light lens with
parameters $I_1/I_s = 37500$, $I_2/I_s = 0$, $z_m \approx 90$. The
region to the right of origin $(x = 0)$ is zoomed and shown in the
inset.

\noindent
the point of the best localization.
Note that the optimized double light mask reduces the atomic background
by a factor of three in the midpoint $(x = 0.25)$ between two deposition
peaks (see the inset of Fig. 6). Moreover, the background in the
optimized double mask scheme is five times smaller compared to the
usual atom deposition in the focal plane (graph not shown) of a single
thick lens.

\section{parameters for optimal squeezing of a thermal atomic beam}
\vskip -0.11in
The effects that have been discussed so far assume an initially
collimated $(\alpha = 0)$ beam of atoms with fixed velocity (or
energy). However, in atom optics experiments involving  thermal
atomic beams, the atoms  possess a wide range of velocities along
the longitudinal (z-axis) and  transverse (x-axis) directions. In
order to characterize the atom spatial squeezing under such
conditions, we need to average the localization factor Eq.
(\ref{local}) over the random initial velocities and angles of the
beam. The averaging can be done by using the normalized
probability density \cite{mcc1}
\begin{equation}
P(v,\alpha) = \frac{1}{2 \sqrt{2\pi}}~\frac{1}{\alpha_o v_o^5}~v^4 \exp\left[- \frac{v^2}{2 v_o^2}
\left(1 + \frac{\alpha^2}{\alpha_o^2}\right)\right]~, \label{aver}
\end{equation}
where $v_o$ is the root mean square speed of atoms with average energy $\bar{E} \equiv m v_o^2/2$.
In the above equation, the term proportional to $v^3 \exp(-v^2/2v_o^2)dv$ represents the thermal
flux probability of having a longitudinal velocity $v$ along the z-direction. The probability of
having a transverse velocity $v_x = \alpha v$ along the x-direction is proportional to the
Gaussian distribution $\exp(-v_x^2/2v_o^2 \alpha_o^2) dv_x$, where $\alpha_o$ is the FWHM of the
angular distribution. Using the probability density ($\ref{aver}$), the averaged localization
factor is thus given by
\begin{eqnarray}
L(z) &=& \frac{2}{\lambda} \int_{\alpha = -\infty}^{\alpha = \infty} \int_{v=0}^{v=\infty}
\int_{x_o = -\lambda/4}^{x_o = \lambda/4} P(v,\alpha) \nonumber \\ 
&&~~~~~\times \left[1 - \cos\left(2 k x(z,x_o)\right) \right] dx_o dv d\alpha \label{mainlocal}~.
\end{eqnarray}
Here, $x(z,x_o)$ represents the solution of differential equations ($\ref{newton}$) for varying
initial conditions $(x_o,v,\alpha)$ at $z \rightarrow -\infty$ of atoms. Note that the solution
of Eq. ($\ref{newton}$) depends on the initial conditions $(v,\alpha)$ through the energy term
$E \equiv m v^2 (1 + \alpha^2)/2$ as well.

Since the focal length of light masks depends on velocity of the
incoming atoms, the velocity spread in the atomic beam leads to
the broadening of the deposited feature size. In the particle
optics context of atom focusing, this is referred to as chromatic
aberration. In addition, the initial angular divergence $(\alpha
\neq 0)$ of the atomic beam degrades greatly the focusing of
atoms. We are interested in the extent to which the velocity and
angular spreads degrade the optimal squeezing of atoms. The best
feature contrast in the presence of aberrations is again defined
by  minimizing the localization factor, Eq. (\ref{mainlocal}). We
have carried out the triple integration in Eq. (\ref{mainlocal})

\begin{figure}[h]
\centerline{ \epsfxsize=210 pt \epsfbox{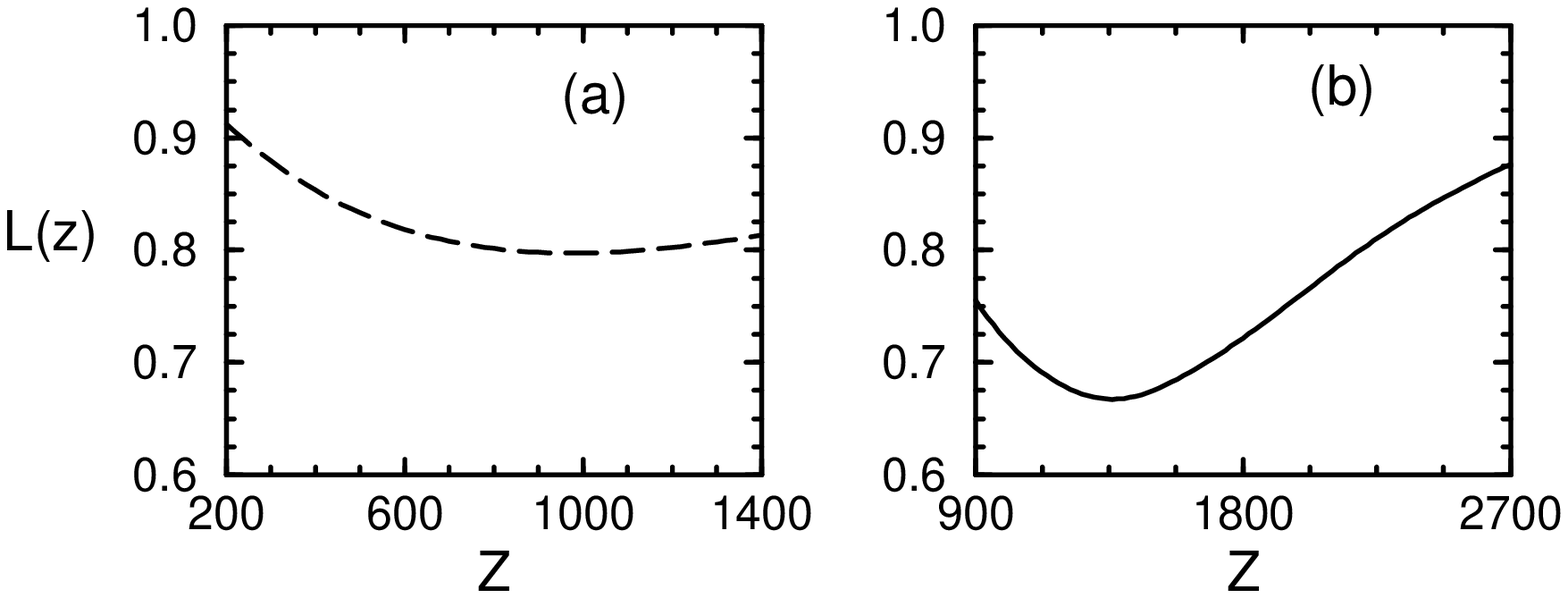}} \centerline{
\epsfxsize=215 pt \epsfbox{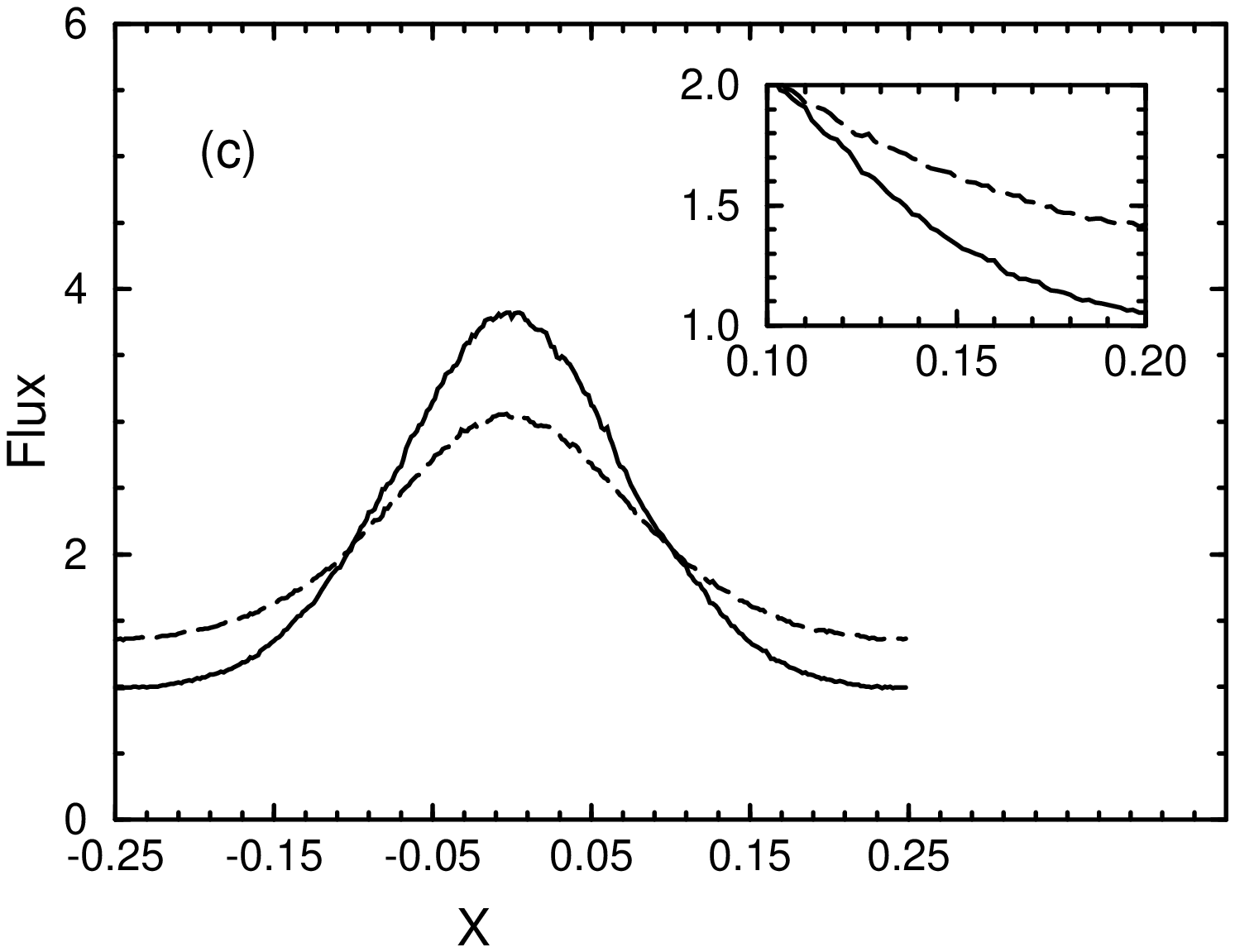}}
\end{figure}
\vskip -0.2in
\noindent
FIG. 7. Localization factor of the atomic distribution for squeezing 
by a single- (a) and double-layer (b) light masks. The parameters used 
are $\bar{E} = 3 \times 10^9$, $\alpha_o = 10^{-4}$, $I_1/I_s = 1500$, 
and (a) $I_2/I_s = 0$, (b) $I_2 = I_1$, $S = S_m \approx 800$. All 
other parameters used are the same as in Table I. The minimal value of
$L(z)$ is 0.67~[0.8] and it occurs at $z = z_m \approx 1350~[975]$
in the graph (b)~[(a)]. The dashed and solid curves in graph (c)
give the atomic distribution at the point $(z_m,S_m)$ of best
squeezing by the single- and double-layer light masks with the
parameters of (a) and (b). The region to the right of origin $(x =
0)$ in graph (c) is zoomed and shown in the inset.

\vskip 0.2in
\noindent
numerically and optimized the localization factor $L(z)$ in the
parameter space $(z,S)$ for the case of the double-layer light
masks. Figures 7 and 8 display  atomic distribution at the point
of the best squeezing by thin-thin and thin-thick lenses
configurations. On comparing the results with those ones for a
single thin or thick lens, it is seen that the thin-thick lens
combination provides the smallest feature size for the atom
deposition. In the case of thin-thin lenses, the effects of
chromatic aberrations are greater because of the strong dependence
of focal length on the atomic velocity. We note that,
though the initial velocity and angular spread of thermal beam
worsen the optimal squeezing of atoms, the effects may become less
important with increasing the number of layers in the multi-layer
light masks. Further, chromatic aberrations can be greatly reduced
by employing low-temperature supersonic beams of highly collimated
atoms.

\section{summary}
\vskip -0.2in
In this paper, we presented the particle-optics analysis for atom
lithography using multiple layers of SW light 
\begin{figure}[h]
\centerline{ \epsfxsize=210 pt \epsfbox{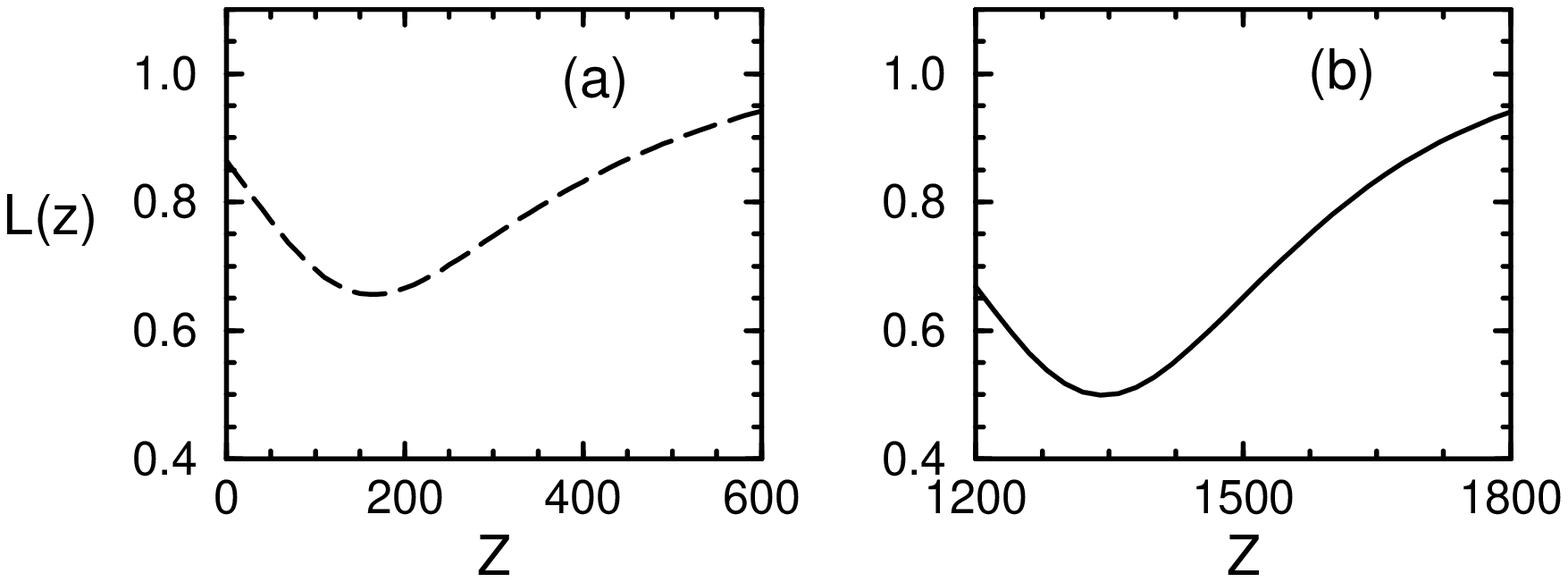}} \centerline{
\epsfxsize=215 pt \epsfbox{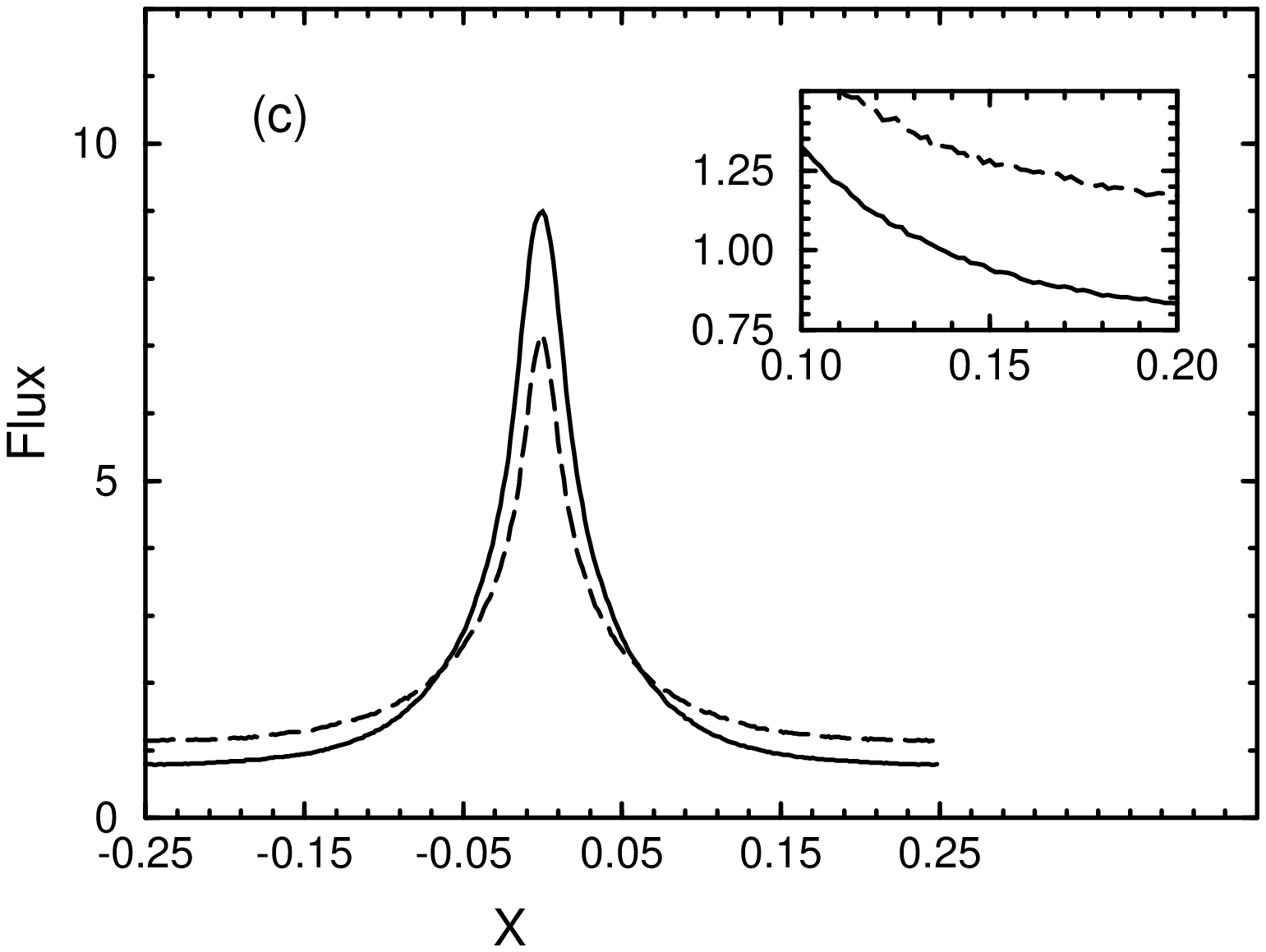}}
\end{figure}
\vskip -0.24in
\noindent
FIG. 8. Localization factor of the atomic distribution for squeezing 
by a single- (a) and double-layer (b) light masks. The parameters used 
are $\bar{E} = 3 \times 10^9$, $\alpha_o = 10^{-4}$, and (a) $I_1/I_s 
= 37500$, $I_2/I_s = 0$, (b) $I_1/I_s = 1500$, $I_2 = 25 I_1$, $S = S_m
\approx 1200$. All other parameters used are the same as in Table
I. The minimal value of $L(z)$ is 0.5~[0.66] and it occurs at $z =
z_m \approx 1350~[160]$ in the graph (b) [(a)]. The dashed and
solid curves in graph (c) give the atomic distribution at the
point $(z_m,S_m)$ of best squeezing by the single- and
double-layer light masks with the parameters of (a) and (b). The
region to the right of origin $(x = 0)$ in graph (c) is zoomed and
shown in the inset.

\vskip 0.2in 
\noindent
masks. In
particular, we studied the spatial squeezing of atoms by a double
layer of standing light waves with particular reference to
minimizing the feature size of atom deposition. At first, linear
focusing of atoms using paraxial approximation was considered.
This showed an improvement in feature contrast at the focal plane,
but the effect was rather modest. We then applied the approach of
optimal squeezing that was suggested recently for the enhanced
localization of atoms in a pulsed SW \cite{mleib}. We showed that
this approach works effectively for atomic nanofabrication and can
considerably reduce the background in the atom deposition.
Based on the optimal squeezing approach, a new figure of merit,
the localization factor, was introduced to characterize the atomic
localization. Both, thin-thin and thin-thick lens regimes of atom
focusing were considered for monoenergetic as well as thermal
beams of atoms. The parameters for the smallest feature size were
found by minimizing the localization factor. We have shown that
using a proper choice of lens parameters, it is possible to narrow
considerably the atomic spatial distribution using the
double-layer light mask instead of the single-layer one. Finally,
we note that our model calculations neglect the effects of atomic 
recoil due to spontaneous emission and the dipole force fluctuations. 
These effects are generally beyond the scope of the classical particle 
optics analysis and can be treated by means of a fully quantum approach. 
A detailed quantum mechanical study of the optimal atomic squeezing in 
application to nanofabrication will be published elsewhere.

\begin{center}
{\bf ACKNOWLEDGMENTS}
\end{center}
This work was supported by German - Israeli Foundation for Scientific
Research and Development.

\end{multicols}

\begin{references}
\bibitem{conserve} We assume the dipole force acting on atoms in the light fields to be
conservative. For the parameters considered in this paper, the velocity dependence of the
light-induced force on atoms can be neglected.
\bibitem{pflux} In all figures, we normalize the flux such that its integral over the
range $[-\lambda/4,\lambda/4]$ has a value of 1.
\bibitem{thin} In the thin-lens approximation, the atomic displacement along the standing wave
axis (x-axis) is negligible within the light fields.
\end{references}
\end{document}